\documentclass[12pt,preprint]{aastex}

\usepackage{epsfig}
\usepackage{graphicx}
\usepackage{epstopdf}
\usepackage{psfig}

\usepackage{natbib}
\begin{document}

\title{A Mid-life crisis? Sudden Changes in Radio and X-Ray Emission from SN 1970G} 
\author{J. A. Dittmann$^{1}$, A. M. Soderberg$^{1}$, L. Chomiuk$^{2}$, R. Margutti$^{1}$, W. M. Goss$^{3}$, D. Milisavljevic$^{1}$, R. A. Chevalier$^{4}$}
\affil{[1] Harvard-Smithsonian Center for Astrophysics, 60 Garden St., Cambridge, MA, 02138; jdittmann@cfa.harvard.edu} 
\affil{[2] Department of Physics and Astronomy, Michigan State University, East Lansing, MI 48824}
\affil{[3] National Radio Astronomy Observatory, Domenici Science Operations Center , Socorro, New Mexico 87801}
\affil{[4] Department of Astronomy, University of Virginia, P.O. Box 400325, Charlottesville, VA 22904-4325}

\begin{abstract}
Supernovae provide a backdrop from which we can probe the end state of stellar evolution in the final years before the progenitor star explodes. As the shock from the supernova expands, the timespan of mass loss history we are able to probe also extends, providing insight to rapid time-scale processes that govern the end state of massive stars. While supernovae transition into remnants on timescales of decades to centuries, observations of this phase are currently limited. Here we present observations of SN\,1970G, serendipitously observed during the monitoring campaign of SN\,2011fe that shares the same host galaxy. Utilizing the new Jansky Very Large Array upgrade and a deep X-ray exposure taken by the Chandra Space Telescope, we are able to recover this middle-aged supernova and distinctly resolve it from the HII cloud with which it is associated.  We find that the flux density of SN\,1970G has changed significantly since it was last observed - the X-ray luminosity has increased by a factor of $\sim 3$, while we observe a significantly lower radio flux of only $27.5 \mu$Jy at 6.75 GHz, a level only detectable through the upgrades now in operation at the Jansky Very Large Array. These changes suggest that SN\,1970G has entered a new stage of evolution towards a supernova remnant, and we may be detecting the turn-on of the pulsar wind nebula. Deep radio observations of additional middle-aged supernovae with the improved radio facilities will provide a statistical census of the delicate transition period between supernova and remnant.

\end{abstract}

\section{Introduction}
Middle-aged radio supernovae span the gap in which the non-thermal radio emission is attributed to shock interaction with the circumstellar material (CSM) as shaped by the evolution of the progenitor star,  and the interstellar material (ISM). This transition occurs on a time-scale of decades and radio studies probe the final stages of stellar evolution including the progenitor star mass loss rate. Several middle-aged radio supernovae have shown strong time evolution in their radio fluxes indicating variable densitiy profiles. For example, historical supernovae in M83 show a wide range of radio emission properties \citep{historical_sne} and observations of SN 1980K from 1994-1995 show a rapid decline in flux (a factor of two below the expected flux from the previous measured power-law decline), which is believed to be due to a decline in the circumstellar density \citep{98k_dropoff}. Similarly, \citet{57D_dropoff} measured a steep decline in radio flux from SN 1957D ($F_{\nu}\propto t^{\alpha}$; $\alpha = -2.9 \pm 0.3$), steeper than other intermediate aged supernovae.

The detection of radio emission from nearby Type IIb supernovae (hydrogen and helium dominate the optical spectra; e.g., \citealt{matheson}) has occurred more frequently with detections of SNe 2001ig \citep{Ryder04}, 2003bg \citep{S06}, 2001gd \citep{gd07}, 1993J \citep{93J_longterm}, 2008ax \citep{roming09}, and 2011dh (\citealt{krauss12,soderberg11dh}), among others, yielding information on the SN shockwave as well as the final stages of massive star evolution. In the future, these objects will illuminate the physical processes that govern the transition from supernova to remnant. Today, however, there are only a handful of known radio supernovae old enough to probe this intermediate regime. 

The early radio emission is typically characterized by a synchrotron spectrum and is generated by interaction of the shock with the CSM  ejected by the star prior to the explosion (\citealt{C82,Chev98}, \citet{CandF06}). In some cases the inferred circumstellar density follows a simple $\rho \propto r^{-2}$ wind profile; however, there is a growing number of radio supernovae that show modulations in their radio light curves. These are usually interpreted to be variations in the circumstellar density (\citet{93J_massloss}, \citet{Ryder04}, \citet{S06},  \citet{Kotak_and_Vink}, \citet{gd07}, \citet{Wellons12}). While it is currently unclear what causes the sudden changes in mass-loss rate, the simplest solution is that the mass loss rate and wind velocity of the progenitor star are not uniform on the timescales we see the radio emission changing, typically representing decades to centuries before the death of the star. Other possible solutions include dynamic instabilities, such as those seen in luminous blue variables like $\eta$ Carinae \citep{etacar_chin} and other instabilities observed in nearby S Doradus type stars \citep{S-Dor_Groot}. Interestingly, some of the mass loss rates inferred from the measured radio emission are larger than the saturation strength for line driven winds, which is around $\dot M \approx 10^{-5} M_\sun$ yr$^{-1}$ \citep{Crow07}. This limit, however, may not apply to lower mass progenitors; standard stellar models prefer smaller mass loss rates \citep{standard_mass_loss_model}, although there is some evidence that the mass loss rates of red supergiants may be larger than these models imply \citep{redgiant_evolution_mass_loss}.

SNe 1993J, 2001gd, and 2004C show evidence for a large mass loss rate and a circumstellar density profile morphologically similar to a uniform shell of material or a power law falling off more shallow than $r^{-2}$, instead of a wind profile (\citet{93J_massloss}, \citet{93J_longterm}, \citet{gd07}). While the wealth of recent detections of supernovae in the radio have provided insight into the final stages of stellar evolution, there are relatively few studied for more than a couple decades and providing information on the final chapter of stellar evolution.  Recovering historical supernovae at radio wavelengths is therefore a rewarding endeavor. 

Here we focus on middle-aged SN 1970G, serendipitously observed as part of our Target-of-Opportunity program for Type Ia SN 20011fe in the same host galaxy, M101 (D = $7.4^{+1.0}_{-1.5}$ Mpc \citet{m101_distance}). SN 1970G was the first supernova detected in the radio, by \citet{Nature_Original_Detection}. After this initial detection, attempts at continuously monitoring SN 1970G were made with the Westerbork Synthesis Radio Telescope (WSRT), and a compilation of these efforts is presented by \citet{Supernova_Last_70s}. Extracting the supernova flux from the flux of the galaxy and the nearby HII region was impaired due to the low resolution of the available radio telescopes \citep{Supernova_Last_70s}. They found that the flux density of SN 1970G was likely below the threshold for detection at 11 cm and 21 cm wavelengths for the first 0.4 years since the optical supernova explosion \citep{Supernova_Last_70s} before ultimately being detected by \citet{Nature_Original_Detection}. SN 1970G remained detectable at approximately $5$ mJy through 1973 when the last marginal detection occurred \citep{Supernova_Last_70s}. Throughout this period, SN 1970G remained undetectable at 6 cm wavelengths below a limit of 1.5 mJy \citep{Supernova_Last_70s}.

In this paper, we present a new detection of SN 1970G using the VLA and the Chandra X-ray Observatory. We detect a point source at the same location of SN 1970G robustly at 6 cm and marginally at 4.5 cm. We demonstrate the increased precision of the Jansky Very Large Array after its recent upgrade that will create new opportunities for studying radio supernovae to late time.  We show that a new emission mechanism in the X-ray regime is now dominating the evolution of the supernova/supernova remnant and attribute it to the likely turn-on of a pulsar wind nebula. We suggest future observations that can confirm the nature of this emission and offer a unique perspective into the formation of pulsar wind nebulae from an observational standpoint.

\section{Observations and Analysis}
\subsection{Karl G. Jansky Very Large Array Observations}
We observed SN 2011fe with the newly updated Karl G. Jansky Very Large Array\footnote{The Karl G. Jansky Very Large
Array is operated by the National Radio Astronomy Observatory, a
facility of the National Science Foundation operated under cooperative
agreement by Associated Universities, Inc.} beginning on 2011 August 25.8 UT, more than $\Delta t = 40$ years after the initial discovery of SN 1970G under program AS1015 (PI: Soderberg), and subsequently  through Director's Discretionary Time program 11B-217 (PI: Soderberg) over six epochs spanning three weeks (see \citealt{ChomiukFE}). Observations were carried out in the A configuration using 2 GHz of bandwidth and recording four polarization products; one baseband of 1 GHz width centered at 5.0 GHz and the other centered at 6.75GHz. Each epoch was 1 hour duration, yielding 40 minutes on source, except for 2011 August 26, when we observed for 2 hours. Data were calibrated using J1349+5341 and 3C286 and were reduced using standard routines in the Astronomical Image Processing System (AIPS). Each baseband was imaged individually; the 6.75 GHz image was then smoothed to the resolution of the 5.0 GHz image, and the two noise-weighted images were averaged together. We created a stacked image by concatenating all UV data for a given baseband and producing a final sky image from the combined dataset. Typical RMS noise for our data is $\approx 3 \mu$Jy per beam, with a beam size of $0.56 \times 0.39$ arcseconds at a position angle of $-79^{\circ}$.

\begin{figure}
\centering
\includegraphics[scale=0.5]{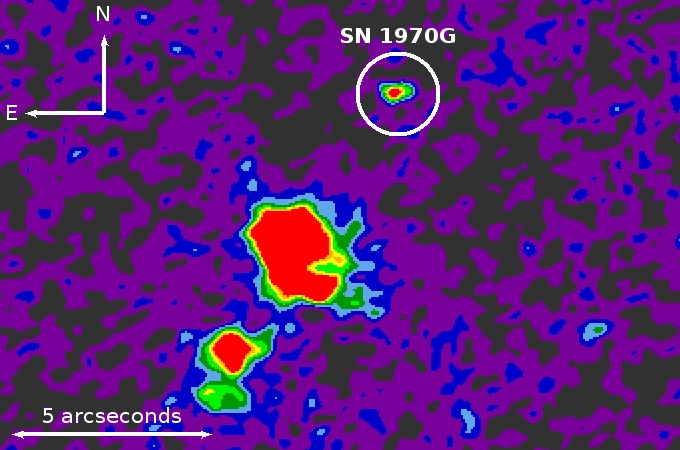} \\
\includegraphics[scale=0.5]{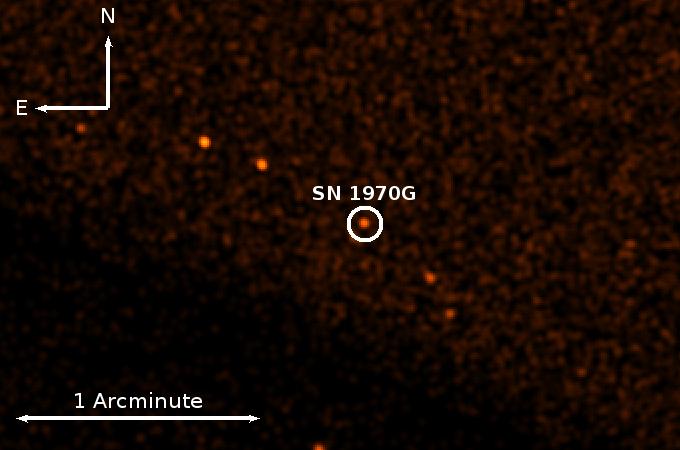} \\
\caption{Top: Jansky Very Large Array data of SN 1970G in M101 (white circle) at a mean frequency of 5.875 GHz taken on 2011 August 25.8. The bright region to the southeast is an HII region associated with the host galaxy. Typical RMS noise is $\sim 3 \mu$Jy per beam, with a beam size of $0.56 \times 0.39$ arcseconds at a position angle of $-79^{\circ}$. Bottom: Image of SN 1970G in M101 from the Chandra X-Ray Observatory taken on 2011 August 27.4. }
\label{images}
\end{figure}

We analyzed the VLA data with the Astronomical Image Processing System (AIPS) by fitting a Gaussian model to the radio supernova in each observation to measure the integrated flux density. We fortuitously discovered a radio point source at RA = 14:03:00.89 $\pm$ 0.01, DEC = 54:14:33.10$\pm$ 0.1, at both 6 cm and 4.5 cm, coincident with the position of SN 1970G. Previous studies have had to proceed carefully to resolve SN 1970G from the nearby HII region NGC 5455 (\citealt{Nature_Original_Detection} ; \citealt{Supernova_Last_70s}), however we note that SN 1970G is clearly resolved and distinct from the nearby HII region in our data (see Figure \ref{images}). The typical noise value in our observations is $3.0 \mu$Jy per beam at 5 GHz and $2.8 \mu$Jy per beam at 6.75 GHz, representing a significant (factor of several tens) improvement from previous data collected by the VLA. We measure a total integrated flux for SN 1970G of $34.9 \pm 4.0$ $\mu$Jy at 5 GHz and $27.5 \pm 5.0$ $\mu$Jy at 6.75 GHz, representing a $\sim 8\sigma$ and $\sim 5\sigma$ detection, respectively. For NGC 5455, we measure a total flux density of $1.8 \pm 0.15$ mJy at 5 GHz and $1.5 \pm 0.15$ mJy at 6.75 GHz, consistent with observations by \citet{70g_1990}. The error in our measurement is dominated by uncertainty in the spatial extent of the region We summarize our flux and astrometric measurements in Table \ref{astrometric_table}

\subsection{Chandra Space Telescope Observations}
We obtained X-ray observations of M101 with the Chandra X-ray Observatory (CXO) on August 27.44 UT under an approved DDT proposal (PI: Hughes) to look for X-Ray emission from SN 2011fe (see \citealt{Raf11fe}). Data were reduced with the CIAO software package (version 4.3), with calibration database CALDB (version 4.4.2). We applied standard filtering using CIAO threads for ACIS data and obtained an effective integration time of 49 ks. A source coincident with the location of SN 1970G was detected in the final image at a significance of $8.3\sigma$ (see Figure \ref{images}).

The Galactic neutral hydrogen absorption in the direction of SN 1970G is $n_{H} = 1.8 \times 10^{20}$ cm$^{-2}$ \citep{m101nh}, derived from 21 cm observations of the Milky Way's HI line.  We correct for the Galactic neutral hydrogen absorption and list unabsorbed fluxes and luminosities in the following way. A circular aperture was placed at the coordinates of the supernova with a size of 
2.5 arcseconds. Assuming an absorbed simple power law spectral model with power-law index $\Gamma=2$ we measure an unabsorbed X-ray flux of $(6.3 \pm 1.8) \times 10^{-15}$ ergs cm$^{-2}$ s$^{-1}$. At a distance of $7.4^{+1.0}_{-1.5}$ Mpc \citep{m101_distance}, this corresponds to an X-ray luminosity of $L_X = (4.1 \pm 1.2) \times 10^{37} $ ergs s$^{-1}$ in the 0.3 to 10 kev band. The same distance and column density of hydrogen from \citet{70g_chandra} were used in order to eliminate systematics between this study and that of \citet{70g_chandra} to investigate the longterm X-ray behavior of SN 1970G . No hard ($E > 4$ kev)X-ray photons were detected in this observation of SN 1970G. There are insufficient photons detected in this observation to investigate whether the spectral index of SN 1970G has changed over these observations.

\section{Discussion}

\subsection{Radio Evolution}
\begin{figure}
\centering
\epsfig{file=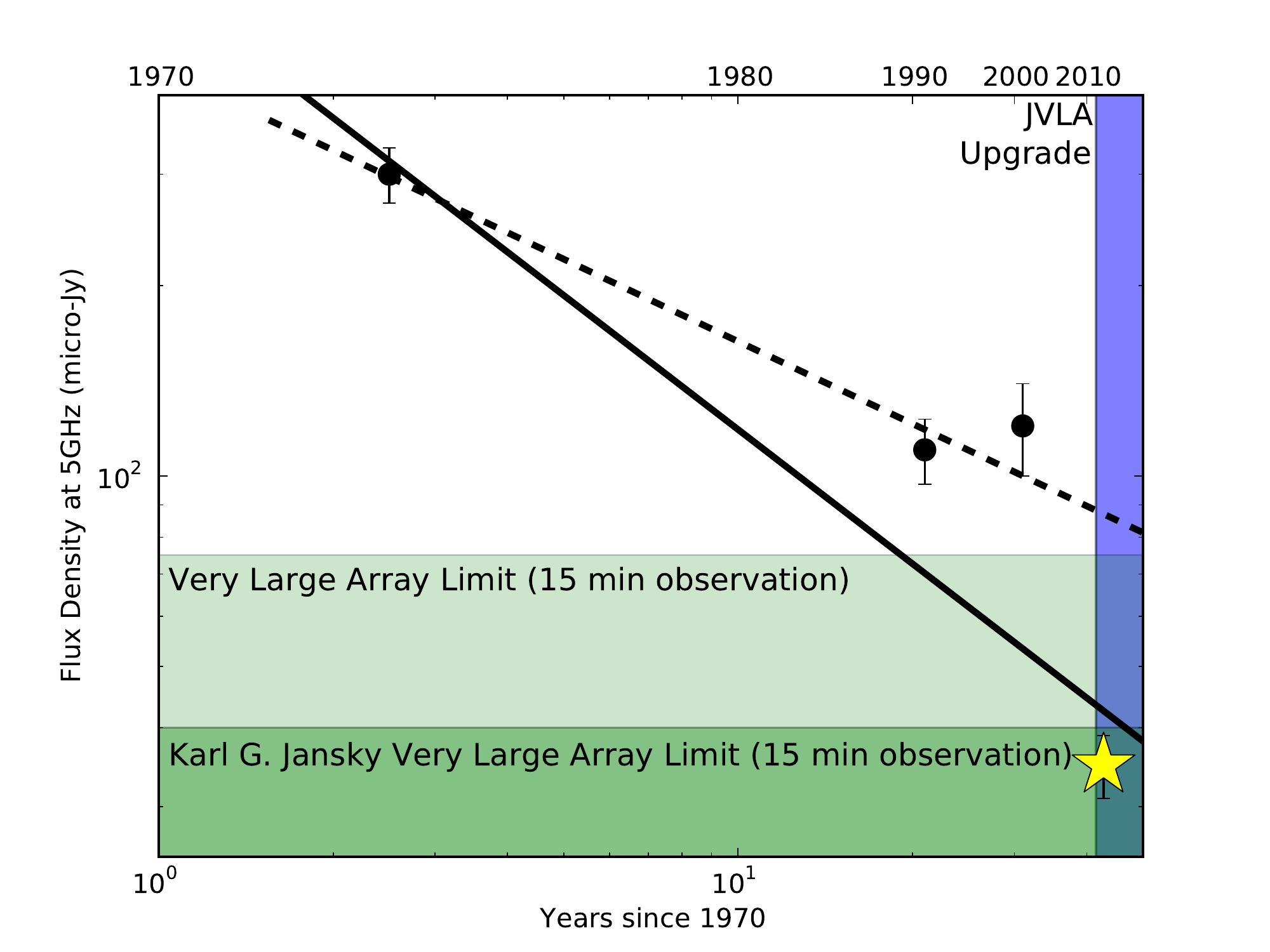,width=0.75\linewidth,clip= } \\
\caption{Radio light curve of SN 1970G over the $\sim 40$ years since explosion at 5 GHz. SN 1970G was not detected during the 1980s, but with only weak upper limits of 0.3 mJy \citep{70g_radio_fade}. We note the increased precision in the Jansky VLA data points during 2011. Between 2000 and 2011 the radio flux of SN 1970G decreased dramatically with an average power law of $F \propto t^{-4.5}$. Historical data come from \citet{Supernova_Last_70s, 70g_1990, 70g_2000}. The 5 GHz flux density for 1990 was obtained by taking the flux density measurements at 20 cm and 3.5 cm reported by \citet{70g_1990} and interpolating to 5 GHz utilizing their reported spectral index.}
\label{70g_radio}
\end{figure}

\citet{70g_2000} measured a flux density of $0.16 \pm 0.02$ mJy at 20 cm and $F = 0.12 \pm 0.02$ mJy at 6 cm between November 2000 and January 2001, far above the flux expected from the trend measured by \citet{70g_1990}, and showing virtually no change from the preceding decade (see Figure \ref{70g_radio}). A flatter spectral index was also measured by \citet{70g_2000} with $F_{\nu}\propto \nu^{\beta}$; $\beta = -0.24 \pm 0.2$ and a temporal index of $\alpha = -0.28 \pm 0.13$ for SN\,1970G \citep{70g_2000}.   Projecting this model forward $\sim$10 years, we expect a flux of 0.10 mJy at the time of our measurement. The current data is able to exclude this model at $\approx 8\sigma$. However, past observations were taken with different configurations of the VLA, with a larger beam size than our observations, and it is possible that there may be some residual contamination in those observations from the nearby HII region, which may be responsible for some of the observed trends. For the rest of this paper, we will assume that the contamination in previous observations is negligible and SN 1970G was distinctly resolved from the nearby HII region. In order to explain the nature of our data, the average power law index in the past decade must have dropped to $\alpha = -4.5 \pm 0.5$. We note that this steepening behavior has also been reported for decades-old supernova SN\,1957D \citep{roger_57d_aas}.  Alternatively, if the radio flux density measured by \citet{70g_2000}  is due to a transient brightening event, then fitting a power law to the remaining data points produces a more modest trend, with the flux density of SN\,1970G decaying as a power law over time with $\alpha = {-0.7 \pm 0.2}$. A time series plot of all radio observations of SN\,1970G at 5 GHz taken to date is shown in Figure \ref{70g_radio}.  At the present rate, the SN will once again fade from radio detectability in the near future.

The drop off in radio luminosity indicates that the forward shock is no longer plowing into a high density CSM and may have entered the local interstellar medium, no longer dominated by interaction with material shed from the progenitor star.  Alternatively, the progenitor star may have been in a quiet, low activity phase prior to increasing in activity in the final decades to centuries before the explosion, depending on the wind velocity of the star at the end stage of its life. Regardless, a steep drop in flux over the past decade indicates that there was a sharp transition in progenitor activity level, lasting $\sim 10^{2} - 10^{3}$ years. It is unclear whether such rapid changes in stellar behavior are driven by the rapid changes in the core of the star, from the stellar envelope, or by interaction with a binary companion. The timescale for changes in the core to be communicated to the surface suggests that binary interactions are a likely explanation for shaping the material far from the system's barycenter. 

\begin{figure}
\centering
\epsfig{file=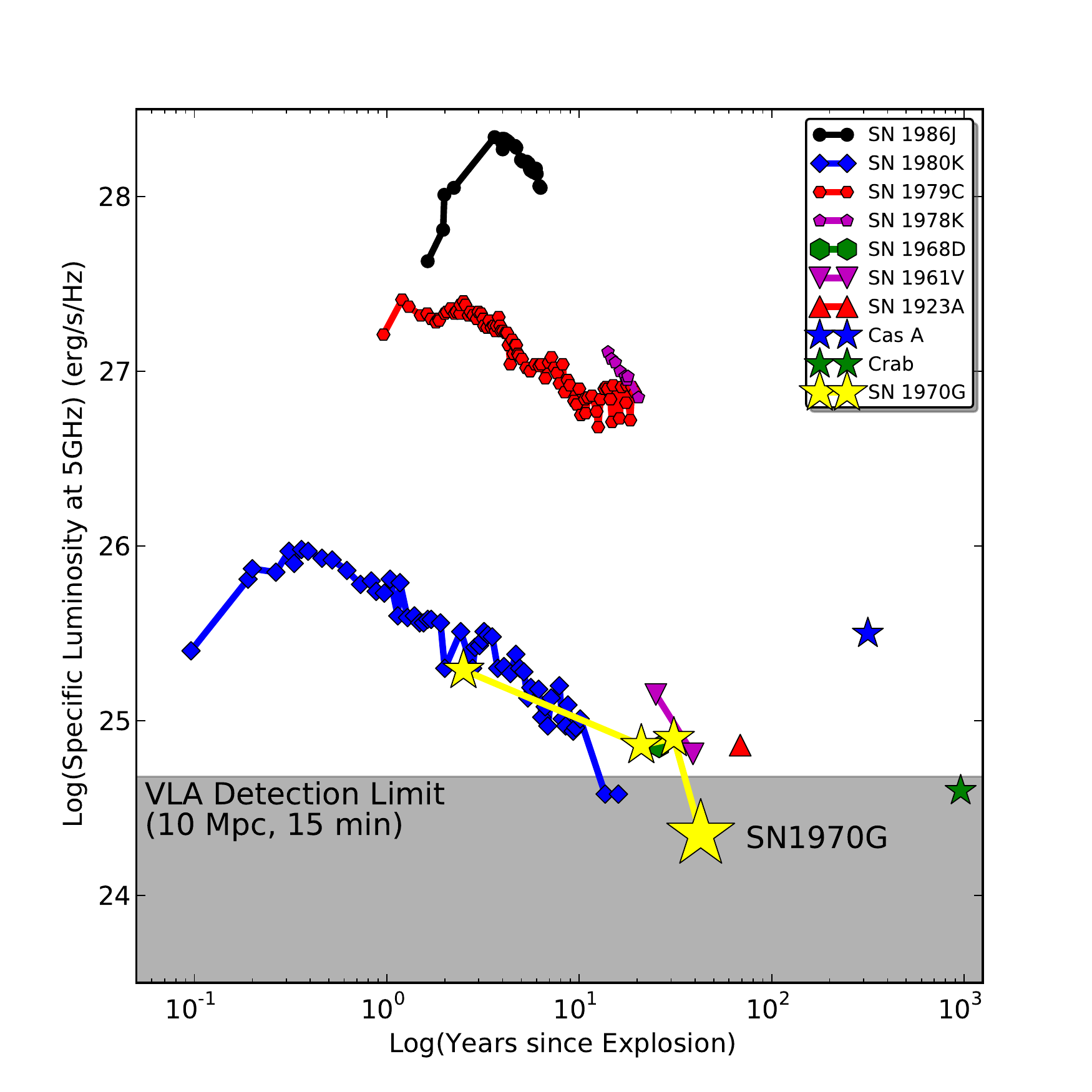,width=0.75\linewidth,clip= } \\
\caption{Radio light curves at 5 GHz of historical supernovae over 25 years old. Middle-aged supernovae vary greatly in flux density, up to several hundreds between different events. We show the detection sensitivity for the new JVLA for an object at 10 Mpc in grey. Our observations of SN 1970G are probing a new regime in the mass loss history of the end-of-life high mass stars, as well as probing the regime bridging the supernovae (left) with the supernova remnants (right).}
\label{radio_all}
\end{figure}

In Figure \ref{radio_all}, we show the radio lightcurve of SN\,1970G compared with other historical supernovae that have been detected in radio and are over 25 years old. The observed radio flux density for these objects can span several orders of magnitude, suggesting a wide diversity in type II SNe, high mass star evolution, and astrophysical environments into which the SNe explode. We find that SN\,1970G is currently the least luminous detected radio supernova, less luminous than the significantly older SN\,1923A. SN\,1970G is beginning to bridge the gap between supernova and supernova remnant, and continued monitoring of this object will illuminate this transition.

\subsection{X-ray Evolution}
\begin{figure}
\centering
\epsfig{file=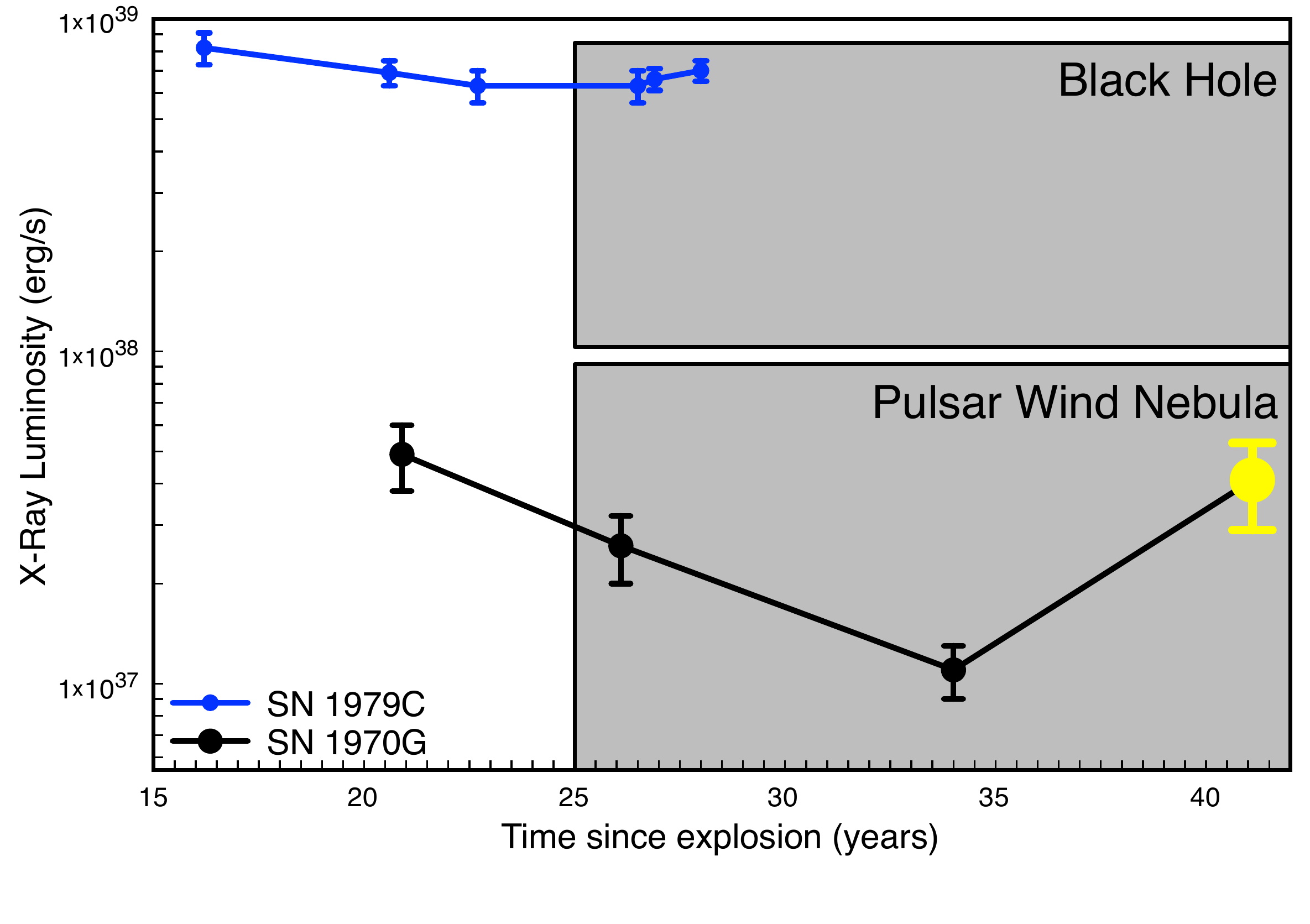,width=0.75\linewidth,clip= } \\
\caption{X-ray lightcurves of SNe 1970G (black) and 1979C (blue) \citet{79C_blackhole} in the $0.3$ - $10$ kev band. We note that both supernovae have longterm variability in their X-ray fluxes and remain detectable decades after the optical supernova has faded from detection. Both supernovae have recently experienced re-brightening in the X-ray, possibly indicative of energy injection by a central engine, such as a pulsar wind nebula (grey swath from \citet{Chev_pulse}) or blackhole accretion (grey swath from \citet{scp+10}).}
\label{1970G_Xray_timeseries}
\end{figure}

Previous Chandra data taken in 2004 by \citet{70g_chandra} indicate SN 1970G has a luminosity of $1.1 \pm 0.2 \times 10^{37}$ ergs s$^{-1}$ in the 0.3-2 keV band with a temporal power law index $1.7 \pm 0.6$, using archival ROSAT and XMM-Newton data. In 2004 there was no evidence for emission above 2 keV, and the spectrum was well described by a thermal model with a temperature of 0.6 keV \citep{70g_chandra}. Projecting this model forward to our observation time, an X-ray flux of $8.7 \times 10^{36}$  ergs s$^{-1}$ is expected in our energy band. However, the luminosity of SN 1970G has increased over this time period. Another middle-age type IIL supernova, SN 1979C, was also recently found to show an increase in X-ray luminosity, interpreted as emission from a central engine, possibly an accreting blackhole \citep{79C_blackhole}. We compare the time series X-ray lightcurves for both of these supernovae in Figure \ref{1970G_Xray_timeseries}. The low number of counts prevents us from determining if the X-ray spectrum of SN 1970G has changed significantly from previous epochs.

It is unlikely that the X-ray emission from SN 1970G is due to accretion onto a stellar mass blackhole. Emission at the Eddington limit, $L_{Edd} = 1.4 \times 10^{38} \left(\frac{M_{BH}}{M_\sun}\right)$ erg s$^{-1}$, would imply an unusually small black hole mass of $M_{BH} \sim 0.1 M_\sun$. Furthermore, there are no $ E > 4$ kev X-ray photons in our detection, which would be expected from accretion onto a black hole. We therefore suggest that the enhanced X-ray emission for SN 1970G is more likely due to the birth of a new pulsar wind nebula than to accretion onto a blackhole, as suggested for SN 1979C \citep{79C_blackhole}. A deeper X-ray exposure of SN 1970G is needed to measure a spectrum and investigate the nature of the X-ray emission. \citet{70g_chandra} characterize the reverse shocked emission as dominated by thermal emission at 0.6 keV while pulsar wind nebulae generally are described with a power law model with $\Gamma \approx 2$ \citep{PWN_review}. Additionally, pulsar wind nebulae can be characterized by a power law spectrum in the radio with a flatter index of $\beta \approx -0.3$ \citep{PWN_review}. Therefore, a multi-wavelength study of the radio and X-ray emission can, in the future, unambiguously distinguish the observational signatures of a newly born pulsar wind nebula from a typical shock interaction model, offering new insight into the formation of these objects from their supernova progenitors. The spectral index between the X-ray and radio regimes can be expected to change if the emission mechanism for SN 1970G has changed. However, current measurement errors in the flux are $\sim 25\%$ in the X-ray and $\sim 10\%$ in the radio band, and its uncertainty prevents us from a robust measurement of a spectral index variation in these bands. We note, however, that we find a relatively flat spectral index between the X-ray and radio, with $\beta \approx -0.5$. Future observations in the X-ray have the potential to address this issue.

\subsection{Interpretation}
We have measured a radio flux density that is nearly an order of magnitude below that expected from observations a decade ago, while simultaneously measuring an X-ray luminosity that has grown by a factor of 4 in the past 4 years. SN 1970G is likely to be entering another stage of evolution on its way to becoming a supernova remnant. Optically, SN 1970G has evolved very little over the past decade \citep{danny_optical}. The strength of the H$\alpha$ line in 2010 was measured as $17 \pm 0.4 \times 10^{-15}$ ergs s$^{-1}$ cm$^{-2}$ \citep{danny_optical}, consistent with the value measured almost two decades earlier by \cite{70g_old_halpha}. Similarly, the type IIL supernova 1979C is also still dominated by H$\alpha$ emission \citep{danny_optical}. While H$\alpha$ emission is expected to decline relative to oxygen lines and the velocity width is expected to narrow at late times, this has not been the case for type IIL SNe 1970G and 1979C, indicating that the late time evolution of these supernovae may be dominated by other processes. \citet{danny_optical} suggests that the persistent H$\alpha$ signal is due to reverse shock heating of hydrogen rich ejecta, and that the strength of the line should fade in the very near future, barring some other excitation mechanism.

X-ray emission from the supernova is principally generated by the reverse shock heating material, which then slowly decays as the material cools. \citet{70g_chandra} measured a continually decaying X-ray flux with a power law index of $\alpha = 1.7 \pm 0.6$, likely due to cooling material from the reverse shock region. Therefore, the recent increase in X-ray flux is unlikely to be due to heating from the reverse shock. 

Recently, X-ray emission from the type IIL SN 1979C has been found to be steady over time, prompting \citet{79C_blackhole} to speculate that the X-ray emission is due to steady accretion onto a compact remnant and not due to emission from the reverse shock region. A resurgence in the X-ray emission of SN 1970G combined with the persistence of H$\alpha$ emission from SN 1970G suggests that SN 1970G is transitioning into a supernova remnant and may be powered by energy injection by a central source heating hydrogen rich inner ejecta. However, \citet{PWN_review} suggest that pulsar wind nebulae have X-ray luminosities typically $\sim L_{X} \sim 10^{35}$ ergs s$^{-1}$, with several orders of magnitude in variation observed. While the emission detected from SN 1970G is relatively high compared to what is expected from a pulsar wind nebula, it is a reasonable mechanism for the observed X-ray emission. Furthermore, we may expect a younger pulsar wind nebula to be more energetic, as the pulsar is spinning down. While we cannot definitively say the strange behavior of SN 1970G in the radio, optical, and X-ray is due to the turn-on of a pulsar wind nebula, the data are consistent with this hypothesis, and future investigations may be able to definitively establish the nature of SN 1970G. We stress that more data is required to determine the true nature of the emission of SN 1970G and is necessary to increase the precision in the current measurement.

\section{Conclusions} 
We present new observations of the middle-aged supernova SN\,1970G in the radio with the Very Large Array and in the X-rays with the Chandra X-ray Observatory. We find that the radio flux has decreased significantly since 2000 to a level of $\approx 33 $ $\mu$Jy. This is approximately a factor of 3 smaller than expected from the study of \citet{70g_2000}. Indeed, in the near future, SN\,1970G will again become undetectable even by the Jansky Very Large Array, if the present decay rate continues. This detection represents one of the faintest objects observed by the Jansky Very Large Array and underscores how valuable this instrument will be in future years in understanding the supernova - supernova remnant transition.

Additionally, we have found an enhancement of the X-ray flux from SN\,1970G over the past 5 years, which we believe is unlikely to result from the reverse shock region, previously shown to be decaying. We find that the X-ray luminosity of SN\,1970G is not atypical for pulsar wind nebula and suggests that we may be witnessing the birth of a pulsar wind nebula from a middle-aged supernova. In the future, we expect SN\,1970G to begin brightening in the radio, as the supernova finally evolves to a remnant. Studying this process is now possible with the improved sensitivity of the Very Large Array, and regular monitoring should help record when this process begins. Additionally, SN\,1957D in M83 has recently been reported to show similar behavior to SN\,1970G, with a steepening of the time-domain behavior of the radio flux density to an index of $\alpha \approx -4.0$, as well as a similar X-Ray luminosity of $\approx 10^{37}$ ergs/s \citep{roger_57d_aas}. This indicates that the physical processes producing this behavior may be common to middle-aged type IIL supernovae.

With RMS noise of $\sim$ few $\mu$Jy, studying radio supernova for many decades can become a routine occurrence. With this ability, new insights into the end stages of massive star evolution (up to centuries before the progenitor exploded) can be obtained. This is a new regime of studying the end states of massive stars, for which SN 1970G represents a unique case.

\acknowledgments
We would like to thank the anonymous referee for his or her comments regarding this paper. Their comments have significantly improved the quality of the work. We thank Harvey Tananbaum and Neil Gehrels for making the Chandra observations possible. The National Radio Astronomy Observatory is a facility of the National Science Foundation operated under cooperative agreement by the Associated Universities, Inc.

\bibliography{sn1970g}

\clearpage

\begin{table}
\begin{center}
\caption{Radio and X-ray Measurements of SN 1970G and NGC 5455}
\begin{tabular}{crrr}
\tableline\tableline
Object & Wavelength & Flux / Flux Density & Epoch \\
\tableline
SN 1970G & $6.75$ GHz & $27.5 \pm 5$ $\mu$Jy & 2011 August 25.8 \\
SN 1970G & $5.0$ GHz & $34.9 \pm 4$ $\mu$Jy & 2011 August 25.8\\
SN 1970G & $0.3-10$ keV & $4.1\times 10^{37} \pm 1.2 \times 10^{37}$ ergs s$^{-1}$ & 2011 August 27.4 \\ 
\tableline
NGC 5455 & $6.75$ GHz & $1.5 \pm 0.15$ mJy & 2011 August 25.8\\
NGC 5455 & $5.0$ GHz & $1.8 \pm 0.15$ mJy & 2011 August 25.8\\
\tableline
\end{tabular}
\end{center}
\label{data_table}
\end{table}

\begin{table}
\begin{center}
\caption{Astrometric Properties of SN 1970G and NGC 5455 from 2011 August 25.8}
\begin{tabular}{crrr}
\tableline\tableline
Object & Parameter & Value \\
\tableline
SN 1970G & Right Ascension (J2000) &14:03:00.89 $\pm$ 0.01 \\
SN 1970G & Declination (J2000) & +54:14:33.1 $\pm$ 0.1 \\
\tableline 

\tableline
\end{tabular}
\end{center}
\label{astrometric_table}
\end{table}

\end{document}